\documentclass[aps,prl,reprint,groupedaddress]{revtex4-1}

\usepackage{graphicx}
\usepackage{xspace}
\usepackage{bbm}
\usepackage{tabularx}
\usepackage{placeins}
\usepackage{url}
\usepackage{mathtools}
\usepackage{bm}

\usepackage{amssymb,amsmath}
\newcommand{\vectorspace}[2]{\ensuremath{\mathbb{#1}^{#2}}\xspace}
\newcommand{\set}[1]{\ensuremath{\mathbf{#1}}}

\newcommand{\Bstrut}[1]{\rule[- #1 ex]{0pt}{0pt}}   

\newcommand*\partialline[2]{\par\noindent\raisebox{.8ex}{\makebox[#1]{\hrulefill\hspace{1ex}\raisebox{-.8ex}{#2}\hspace{1ex}\hrulefill}}}

\newcounter{schemeequation}

\makeatletter
\newcommand{\reqnum}{\refstepcounter{schemeequation}\textup{\tagform@{S\theschemeequation}}}
\makeatother

\usepackage{newfloat}
\DeclareFloatingEnvironment[
    fileext=los,
    listname=List of Schemes,
    name=Scheme,
]{mipmodel}

\usepackage{graphicx}

\begin{document}


\title{Rich Ground State Chemical Ordering in Nanoparticles: Exact Solution of a Model for Ag-Au Clusters.}



\author{Peter Mahler Larsen}
\author{Karsten Wedel Jacobsen}
\author{Jakob Schi{\o}tz}
\email{schiotz@fysik.dtu.dk}
\affiliation {Center for Atomic-scale Materials Design (CAMD), Department of Physics, Technical University of Denmark, 2800 Kongens Lyngby, Denmark}

\date{\today}

\begin{abstract}
We show that nanoparticles can have very rich ground state chemical order.
This is illustrated by determining the chemical ordering of 
Ag-Au 309-atom Mackay icosahedral nanoparticles.  The energy of the
nanoparticles is described using a cluster expansion model, and a Mixed
Integer Programming (MIP) approach is used to find the exact ground state
configurations for all stoichiometries.  The chemical ordering varies
widely between the different stoichiometries, and display a rich zoo
of structures with non-trivial ordering.
\end{abstract}

\pacs{}

\maketitle

Ever since the surprising discovery by Haruta \emph{et al.}\ that gold
is catalytically active in nanoparticulate
form~\cite{Haruta:1987Novel}, there has been intense research into the
catalytic properties of gold~\cite{Daniel:2004hx,Ishida:2016fx,Sardar:2009jr,mingshu2006ultrathin,gao2010aucatalysts}
and silver~\cite{Guo:2008kq} nanoparticles, including bimetallic Ag-Au
nanoparticles~\cite{Liu:2005dr,Chang:2008ch,Raveendran:2006cd}.  These particles
also display interesting optical and plasmonic properties, see
e.g.~the reviews by Feng \emph{et al.}~\cite{Feng:2010cx} and Boote
\emph{et al.}~\cite{Boote:2014gw}, and show promising medical
applications~\cite{Bachelet:2016eb}.

The catalytic properties of a nanoparticle often depend critically on
the detailed atomic
configuration~\cite{Honkala:2005iu,Siahrostami:2013he}.  This is
particularly important for bimetallic nanoparticles, which can
preferentially exhibit one or the other material on the surface, and
often can be designed in so-called core-shell structures with one of
the metals as the catalytically active shell.  It has been
demonstrated that the ability to tailor materials that naturally forms
desirable atomic-scale structures may significantly enhance catalytic
activity and/or selectivity towards the desired
reaction~\cite{Siahrostami:2013he}.

It is thus important to be able to predict the shape and chemical
ordering of nanoparticles.~\cite{Ferrando:2008kv,Rossi:2004kv}  This is, however, a difficult task both
due to the difficulty of calculating the energy of a given
configuration accurately, but mostly due to the very large
configurational space that one must sample.  This is usually done with
Monte Carlo based techniques such as genetic
algorithms~\cite{Lysgaard:2015kb,Hartke:1993kw,Rossi:2005jy}, simulated
annealing~\cite{Lee:2003kp}, basin hopping~\cite{Wales1999}, or minima
hopping~\cite{Goedecker:2004kt}.  While all these methods can
efficiently find configurations that are close to the global minimum,
one cannot in principle know how close to the optimum the solutions
are, nor can one know for sure if the global optimum has been found.

In this Letter we address the relatively simple case of bimetallic
Au-Ag nanoparticles with 309 atoms in the Mackay icosahedral form.
This is one of the so-called \emph{magic number} structures where the
per-atom energy is particularly low, leading to a morphology that is
very robust to changes in stoichiometry and chemical ordering.  We
therefore only consider the chemical ordering, keeping the morphology
constant.  Even in this case, the search space is so large that it is
unlikely that the stochastic methods find the ground state, for
example there are $4.7\cdot 10^{91}$ possible chemical orderings of
the $\text{Ag}_{154}\text{Au}_{155}$ cluster.  We address this by
describing the energy of the nanoparticle using a Cluster Expansion (CE)
model~\cite{sanchez1984cluster} fitted to the semi-empirical Effective Medium Theory (EMT)~\cite{jacobsen1996semi} potential.  This allows us to use a Mixed Integer Programming (MIP)~\cite{nemhauser1988integer, chen2010applied}
approach to provably find the chemical ordering of the ground state
configuration.
Details about the method can be found in the section on the computational approach.

\begin{figure*}[t]
\centering
\begin{tabular*}{\textwidth}{cccc}
\includegraphics[width=0.20\textwidth]{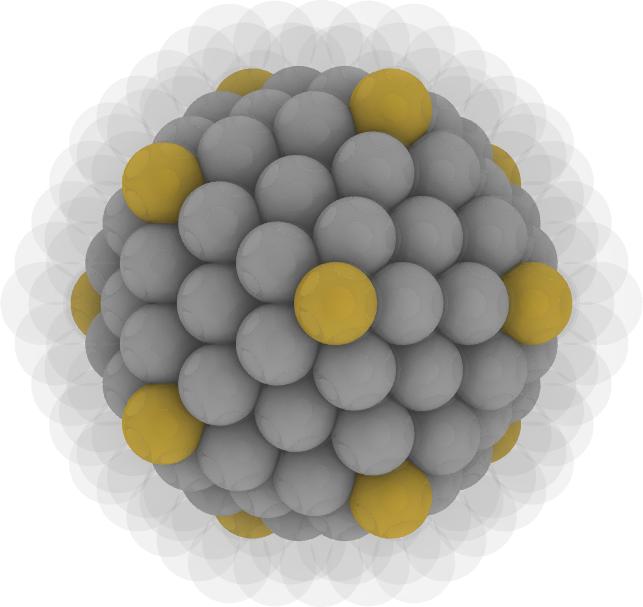} &\hspace{6mm}
\includegraphics[width=0.20\textwidth]{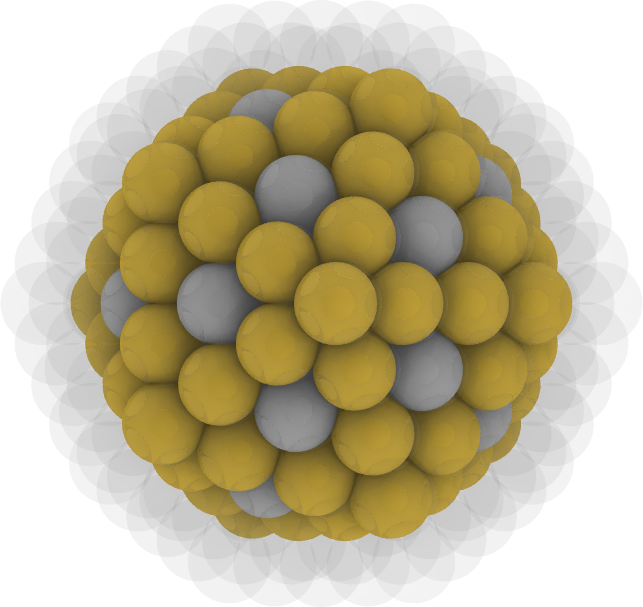} &\hspace{6mm}
\includegraphics[width=0.20\textwidth]{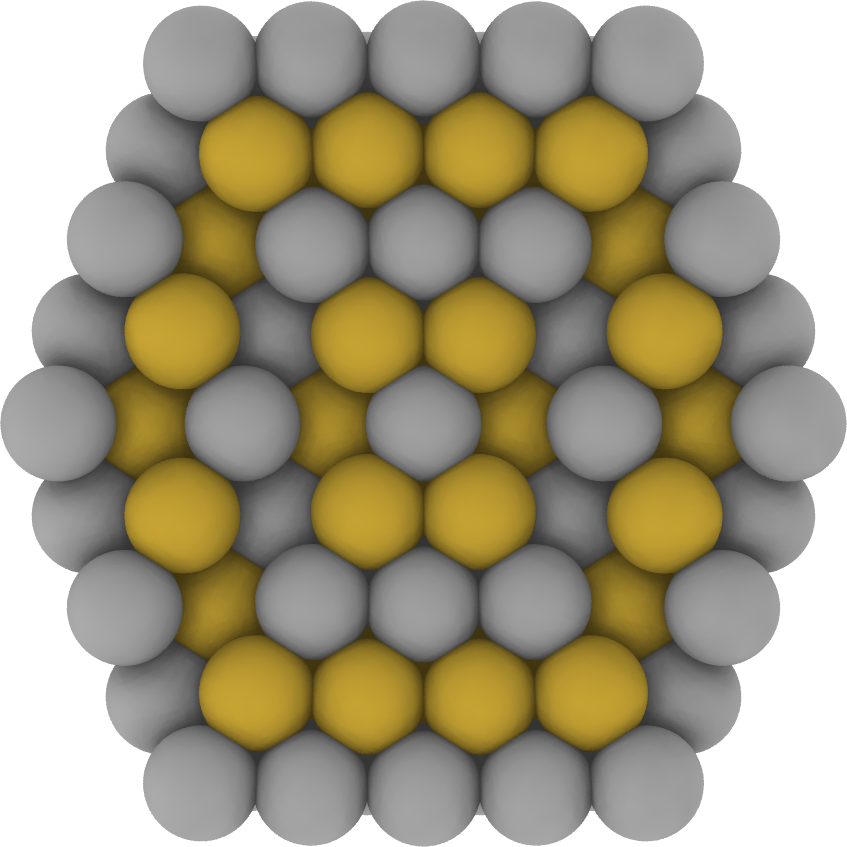} &\hspace{6mm}
\includegraphics[width=0.20\textwidth]{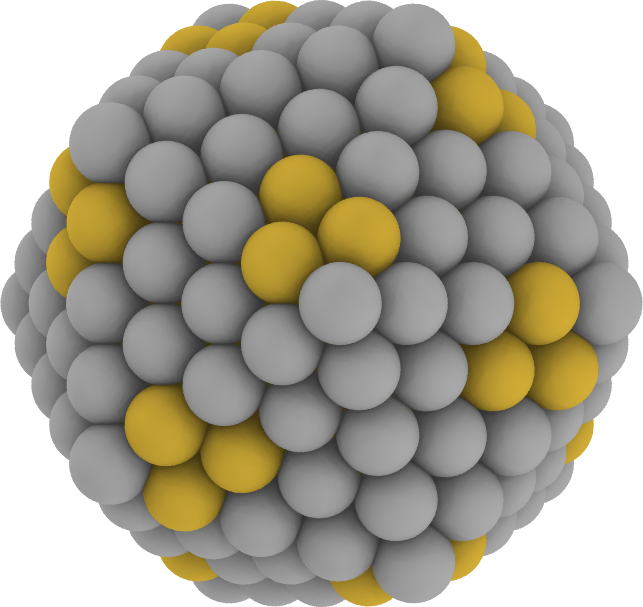}\\
\textbf{(a)} $\textbf{Ag}_{\textbf{296}}\textbf{Au}_{\textbf{13}}$ &
\textbf{(b)} $\textbf{Ag}_{\textbf{228}}\textbf{Au}_{\textbf{81}}$ &
\textbf{(c)} $\textbf{Ag}_{\textbf{205}}\textbf{Au}_{\textbf{104}}$ &
\textbf{(d)} $\textbf{Ag}_{\textbf{162}}\textbf{Au}_{\textbf{147}}$
\vspace{3mm}\\

\includegraphics[width=0.20\textwidth]{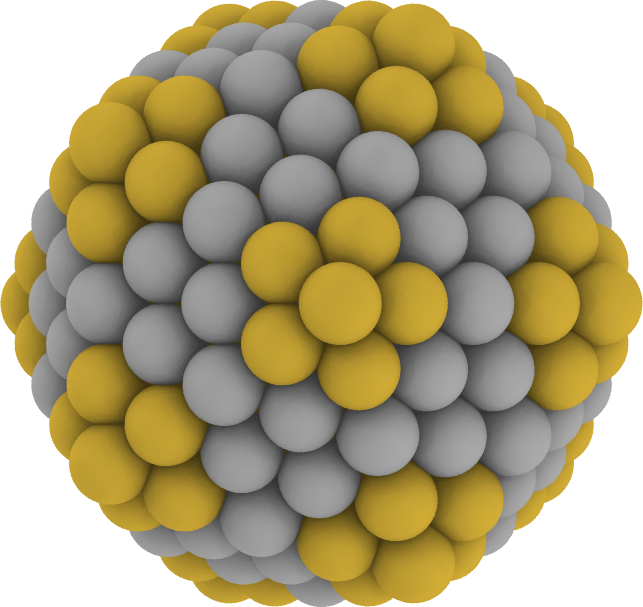} &\hspace{6mm}
\includegraphics[width=0.20\textwidth]{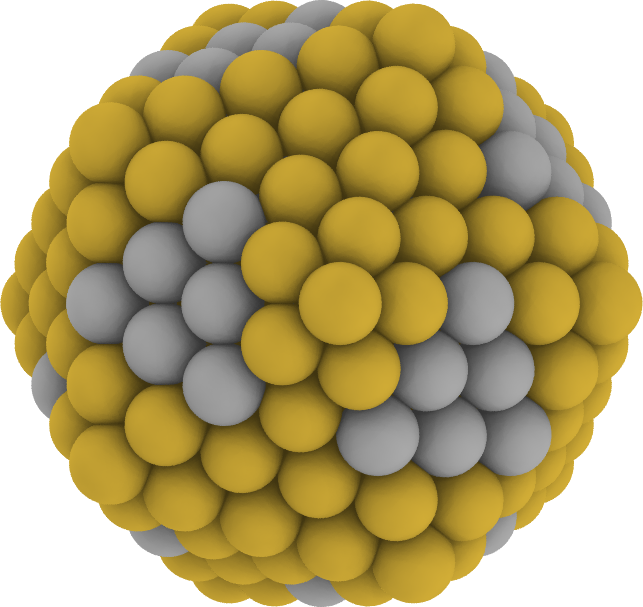} &\hspace{6mm}
\includegraphics[width=0.20\textwidth]{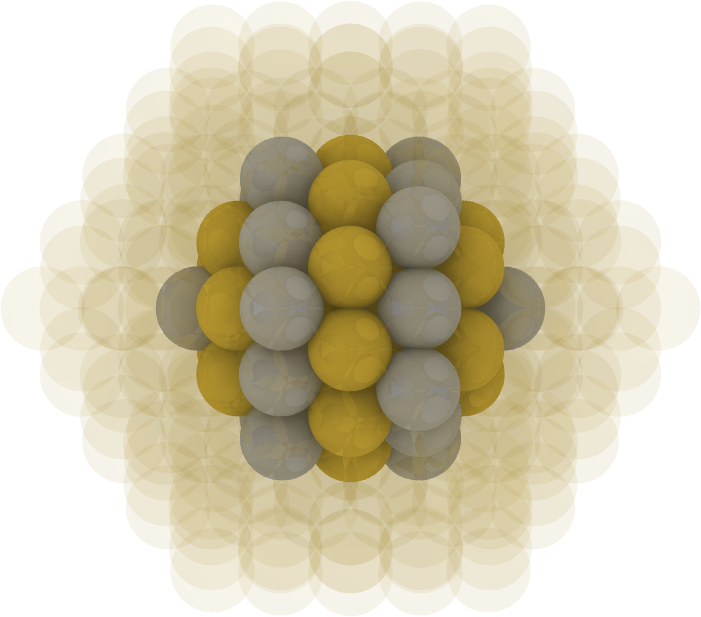} &\hspace{6mm}
\includegraphics[width=0.20\textwidth]{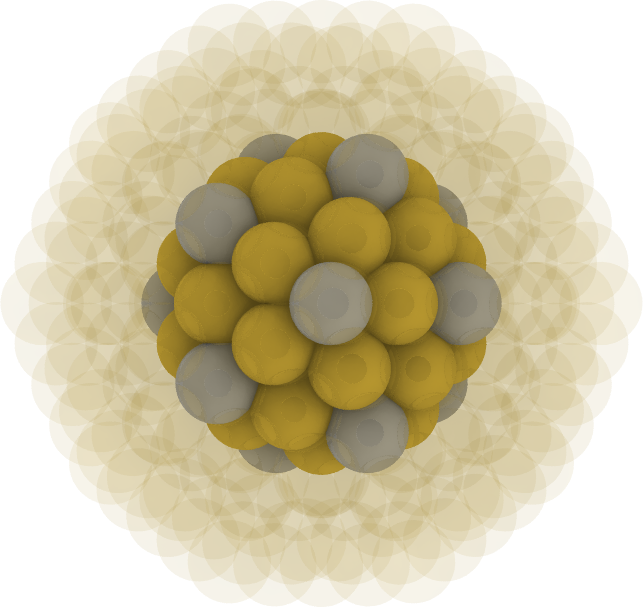}\\
\textbf{(e)} $\textbf{Ag}_{\textbf{126}}\textbf{Au}_{\textbf{183}}$ &
\textbf{(f)} $\textbf{Ag}_{\textbf{75}}\textbf{Au}_{\textbf{234}}$ &
\textbf{(g)} $\textbf{Ag}_{\textbf{25}}\textbf{Au}_{\textbf{284}}$ &
\textbf{(h)} $\textbf{Ag}_{\textbf{13}}\textbf{Au}_{\textbf{296}}$
\end{tabular*}
\caption{A selection of ground-state configurations in a 309-atom Mackay icosahedron Ag-Au nanoparticle.  The configurations exhibit a diverse range of highly ordered interior 
and surface 
structures.  In \textbf{(a)} and \textbf{(b)}, the transparent atoms represent a single layer of Ag atoms; in \textbf{(g)} and \textbf{(h)} they represent two layers of Au atoms.
The Au atoms exhibit a strong preference for subsurface shell sites, in particular the subsurface corners \textbf{(a)}, followed by the subsurface edges \textbf{(b)}.  The Ag atoms show a strong preference for $2^\text{nd}$ shell corner sites \textbf{(g)}-\textbf{(h)}, as well as the central atom.  The lowest energy configuration \textbf{(c)} (shown as a slice through the nanoparticle) has a perfectly ordered onion-shell structure.  At high Au concentrations, the preference of Au atoms for corner, edge or interior sites of the surface facets varies as a function of composition \textbf{(d)}-\textbf{(f)}.
\label{fig:renders}}
\end{figure*}

\begin{figure}[t]
\centering
  \includegraphics[width=\linewidth]{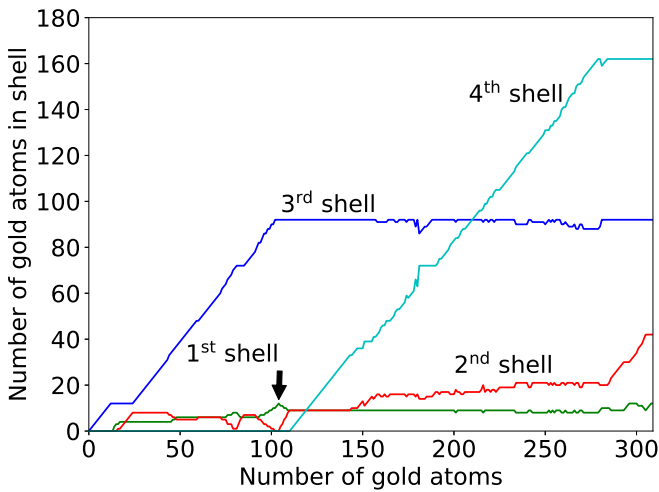}
  \caption{Number of Au atoms in each shell vs. the total number of Au atoms in the nanoparticle.  The structural evolution exhibits a complex interplay between the shells, with no monotonically increasing shell occupancies.
  \label{fig:shell_counts}}
\end{figure}

\begin{figure}[t]
  \centering
  \includegraphics[width=\linewidth]{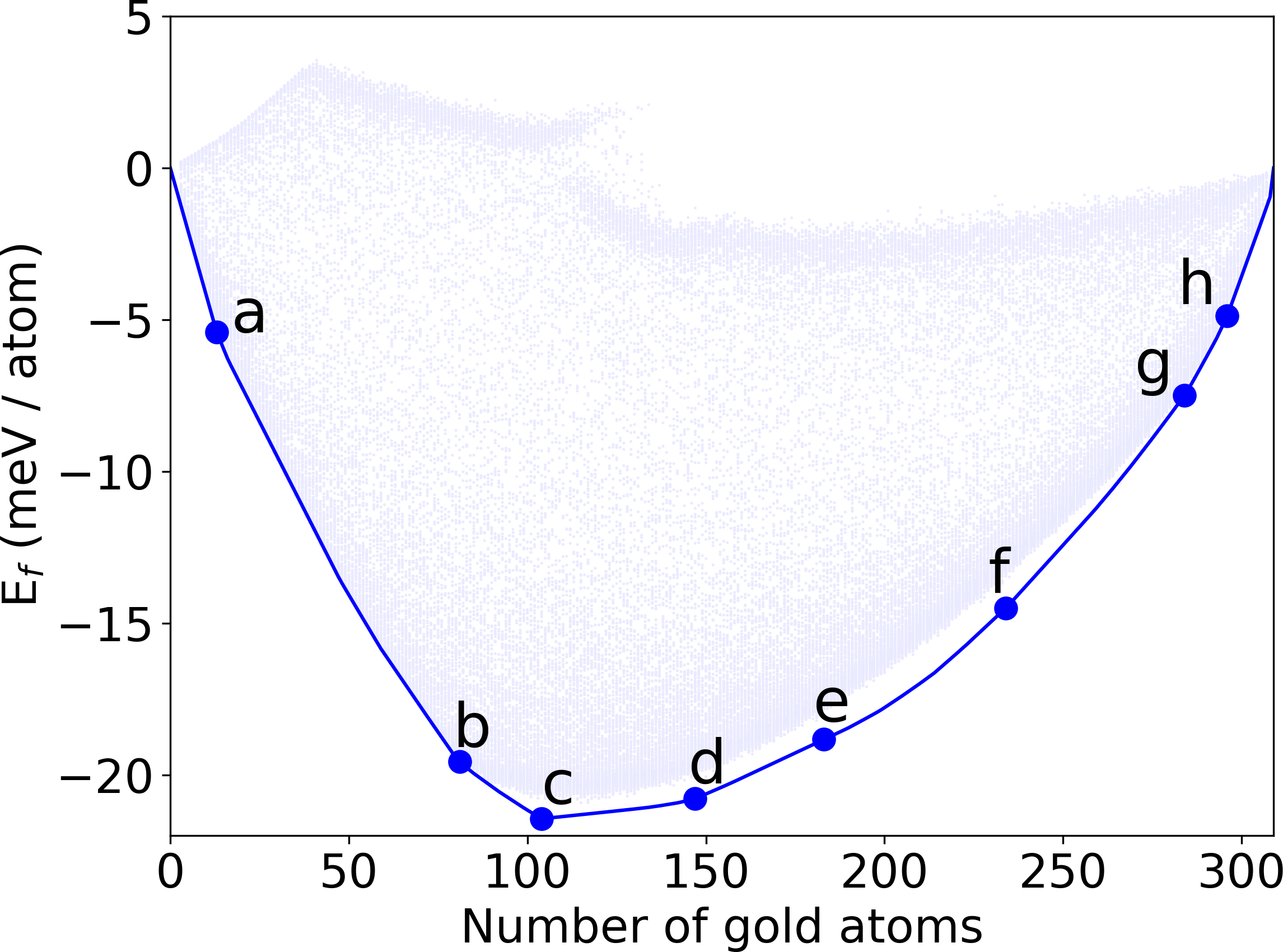}
  \caption{Energy of formation (from EMT) vs. the number of Au atoms in the nanoparticle.  The grey dots show configurations sampled for CE model construction.  The blue line shows the convex hull of the ground state configurations.  Compositions at which the nanoparticle exhibits strong ordering (shown in Figure~\ref{fig:renders}) are marked with a circle.
  \label{fig:cluster_hull}}
\end{figure}

We find ground state configurations displaying a rich zoo of
ordered structures, where the sites exposed on the nanoparticle
surface vary dramatically with varying chemical composition of the
nanoparticle.
The configurations are not intuitive ground states,
fitting into none of the well-studied categories of
core-shell, Janus, or phase mixing nanoparticles~\cite{Wang:2013hu}.

\section*{Results}
We have minimized the energy to find the optimal configuration of the nanoparticle at every composition between 0-309 Au atoms. The nanoparticle exhibits a surprisingly complex structural evolution, as shown in Figure~\ref{fig:renders}.  Figure~\ref{fig:shell_counts} shows the number of Au atoms in each shell as a function of composition.  Quite noticable is that no shell shows a monotonic increase in Au content.  Here, the second shell is particularly illustrative, in that it increases in Au content from 18 Au atoms, but has zero Au content again at 104 Au atoms.  A further example (not shown in Figure~\ref{fig:shell_counts}) is the central atom, which is occupied by a Au atom at 13 and 309 Au atoms only.

Figure~\ref{fig:cluster_hull} shows the convex hull of the formation energy of the ground-state configurations at every composition.
The formation energy of a configuration $\bm{\sigma}$
with fractional compositions $x_{\text{Ag}}$ and $x_{\text{Au}}$ is given by:
\begin{equation}
E_f(\bm{\sigma}) = \frac{1}{309}\left[ E(\bm{\sigma}) - x_{\text{Ag}}E_{\text{Ag}} - x_{\text{Au}}E_{\text{Au}} \right]
\label{eq:formation_energy}
\end{equation}
The convex hull contains a high number of compositions (99), which causes it to be a largely smooth function of composition.  As such, the compositions highlighted are those where the energy gradient changes rapidly.  These configurations (shown in Figure~\ref{fig:renders}) are remarkable in that they exhibit strong ordering, either rotationally symmetric ordering (Figures~\ref{fig:renders}a-c, g-h), or ordered geometric patterns (Figures~\ref{fig:renders}d-f).


\begin{figure*}[t]
\centering
\begin{tabular*}{\textwidth}{cc}
\partialline{0.47\linewidth}{\textbf{Ag atom in Au nanoparticle}} &\hspace{1mm}
\partialline{0.47\linewidth}{\textbf{Au atom in Ag nanoparticle}}\vspace{1mm}\\
\includegraphics[width=0.49\textwidth]{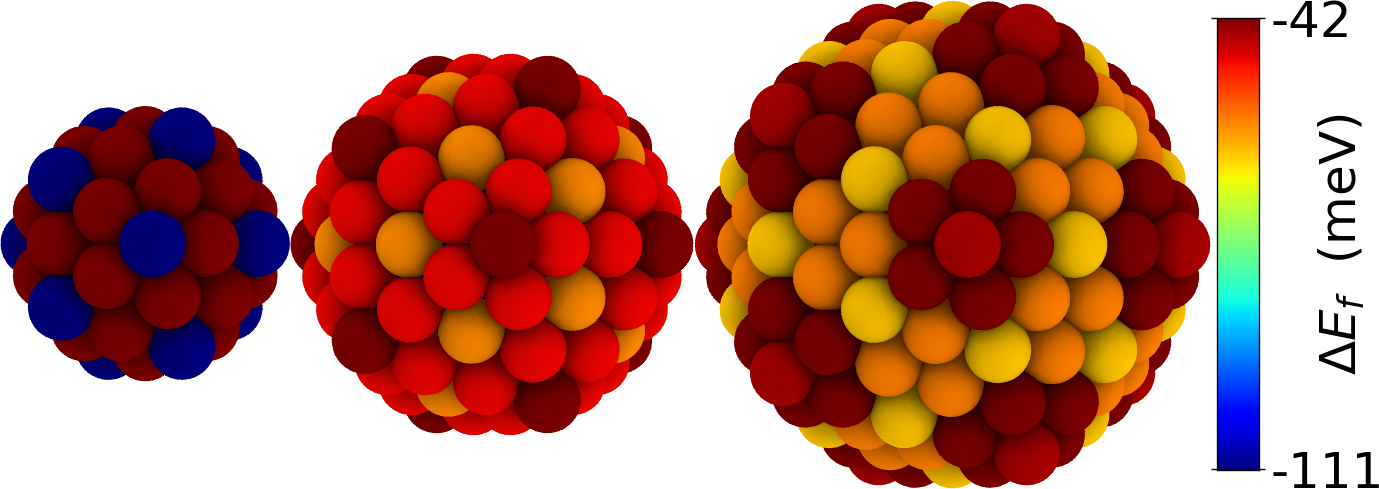} &\hspace{1mm}
\includegraphics[width=0.49\textwidth]{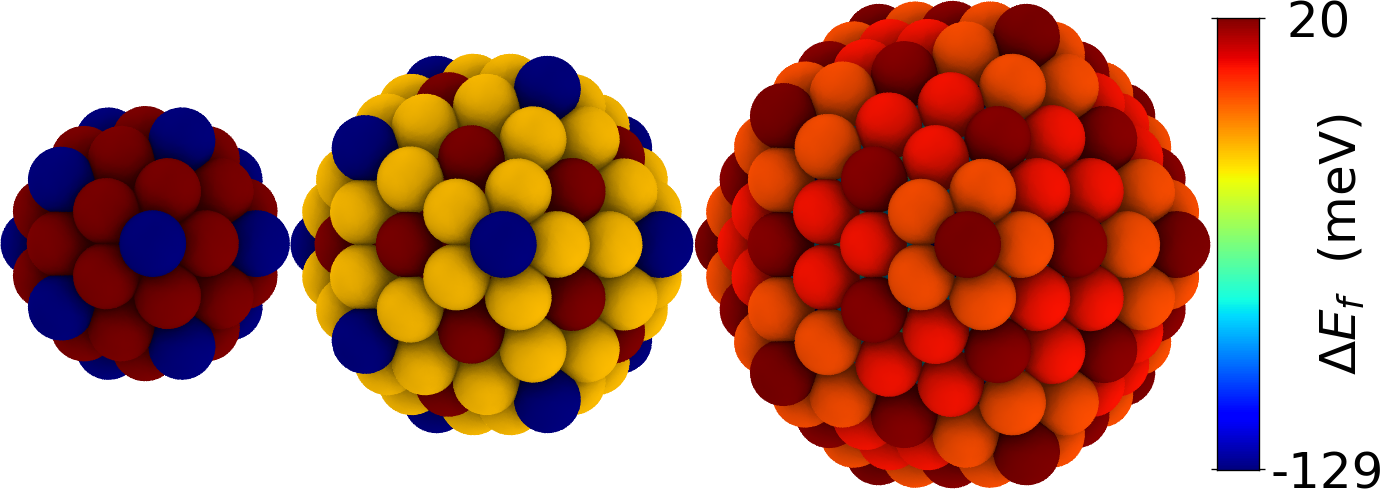}
\end{tabular*}
\caption{Change in formation energy in pure nanoparticles upon the replacement of a single atom with the opposite
species, for all sites in the outer three shells.  This simple measure accounts for much of the variation in the ground state structures, but is tempered at higher concentrations by a preference for heteroatomic bonding.
\label{fig:site_energies}}
\end{figure*}

In many cases, regular patterns are formed inside the nanoparticle.  Of particular interest is the Ag$_{205}$Au$_{104}$ cluster shown in Figure~\ref{fig:renders}c, which forms an onion-like structure with alternating layers of pure Ag and Au.  This structure has perfect icosahedral symmetry and is also the cluster with the most negative formation energy.  A similar 5-ring onion-structure has been proposed by Cheng \emph{et al.}~\cite{cheng2006onion} for Pd-Pt clusters, also using a semi-empirical potential.
The symmetric configurations at Ag$_{296}$Au$_{13}$ and Ag$_{13}$Au$_{296}$ (shown in Figures~1a and 1h respectively) highlight the preference for occupation of subsurface corners for Au atoms, and $2^{\text{nd}}$ shell corners for Ag atoms.

The flower-like surface decoration shown in Figure~\ref{fig:renders}e is energetically favourable in the local concentration region.  As can be seen in Figure~\ref{fig:shell_counts}, the number of Au atoms in the $4^{\text{th}}$ shell increases almost linearly in the range 110-284 Au atoms.  However, there is a noticeable plateau in the region 181-190 Au atoms, where the flowers are present.  Although the surface layer of the flower-structure has icosahedral symmetry, the symmetry is broken in the interior layers.

From a graph theoretical perspective, the triangular Ag islands in the surface layer of the structure at 234 Au atoms (shown in Figure~\ref{fig:renders}f) exhibit a further peculiarity.  Although the symmetry is broken (and one island is missing an atom), the 8 islands fulfil the criteria of a maximum independent vertex set~\cite{bondy1976graph} in the dodecahedral platonic graph.  This graph consists of 20 vertices (one for each facet of the nanoparticle) and 30 edges, which exist between adjacent facets.  This means that no more than 8 islands can be placed on the surface without overlap.

Other than strong geometric ordering, a characteristic of the structures is the change in preference for occupation of different site types as a function of concentration.  For example, the nanoparticle in Figure~\ref{fig:renders}d shows how Au atoms form three-atom islands next to, but not at the corners of the icosahedron.  When the Au content is increased, however, the Au forms islands which switch to being centered on the corners (Figure~\ref{fig:renders}e).  Similarly, Ag atoms disfavour subsurface corners (Figure~\ref{fig:renders}a), but partially occupy these sites when the flower-structures are present in the surface layer; this effect is shown in Figure~\ref{fig:shell_counts} by the dip in $3^{\text{rd}}$ shell occupation which accompanies the plateau in the $4^{\text{th}}$ shell (again, in the region 181-190 Au atoms).

All nanoparticles with up to 111 Au atoms only present Ag atoms in the surface, and all particles with more than 282 Au atoms only present Au atoms in the surface, in spite of the lower surface energy of Ag.  This is contrary to the results of studies of other bimetallic nanoparticles in which the driving force of surface segragation is the relative surface energies~\cite{schoeb1992drivingforce}.
Instead, the driving force \cite{kozlov2015determine} of the structural evolution  as a function
of composition is a trade-off between the large energetic differences
between site types (shown in Figure~\ref{fig:site_energies}) and a
preference for Ag and Au to form heteroatomic bonds, which in bulk
materials at 0K results in the formation of ordered
alloys. See also Supplementary Information section S1.

To a first approximation, ground state configurations are stable at moderate temperatures if the first excited state differs in the site-occupancy or nearest-neighbour coordination statistics.
At a majority of concentrations the density of low-energy states is high.  Nonetheless, we have identified more than 10 ground-state structures where the energy gap to the first excited state is in the range $10-30 \text{meV}$ (see Supplementary Information section S2); this suggests that production and observation of some of these structures should in principle be possible in a moderately cooled laboratory setting.

The remarkably rich chemical ordering has been obtained despite the use of a relatively simple semi-empirical potential, with no angular-dependent terms and a significantly lower complexity than a full ab initio Hamiltonian.
We note that, regardless of the energetic model used, Monte Carlo methods are incapable of determining the optimality of a configuration; the conclusions of the present work rely on the ability of the MIP model to guarantee that the structures found are indeed the ground state structures of the CE model.

The nanoparticle ground-state structures can viewed online at the Computational Materials Repository~\cite{cmr_database}.

\section*{Computational Approach}

Given a fixed site geometry, a cluster expansion uses pseudo-spin variables at each site in conjunction with an orthogonal basis (the clusters) to model configurational properties of the system.  Specifically, a cluster Hamiltonian is of the form:
\begin{equation}
\begin{aligned}
E(\bm{\sigma}) = &V_0
+ \sum\limits_{i \in \set{C}_1} V_{i}^{(1)} \sigma_i
+ \sum\limits_{(i,j) \in \set{C}_2} V_{i,j}^{(2)} \sigma_i \sigma_j\\
&+ \sum\limits_{(i,j,k) \in \set{C}_3} V_{(i,j,k)}^{(3)} \sigma_i \sigma_j \sigma_k
+ \ldots
\end{aligned}
\label{eq:cluster_hamiltonian}
\end{equation}
where $E(\bm{\sigma})$ is the energy of a configuration $\bm{\sigma}$,
$\set{C}_n$ is the set of all $n$-body clusters, each containing
cluster instances $i$, $(i,j)$ or $(i,j,k)$ for 1, 2 or 3-atom
clusters respectively, collectively referred to as $\set{c}_f$ in the following; $V_{c_f}^{(n)}$ are the effective cluster interactions (ECI) for the $n$-body cluster instances; and $\sigma_i$ is the pseudo-spin variable at each site $i$.

A standard transformation is to change the spin variables $\sigma_i
\in \{-1, 1 \}$ to binary variables $x_i \in \{0, 1\}$ using the
relation $\sigma_i = \left( 2x_i-1 \right)$ which produces an
equivalent Hamiltonian with different ECIs but only binary variables.
It can be written compactly as 
\begin{equation}
  E(\bm{x}) = E_0 +
  \sum_{\set{c}_f} E_{\set{c}_f} \prod_{i \in \set{c}_f}
  x_i,
\end{equation}
where $E_{\set{c}_f}$ are the new ECIs.
With a Hamiltonian in this form, we can formulate a MIP model in order to find provably optimal configurations.  MIP models, which are a generalization of linear programming models, solve problems of the form:
\begin{center}
\begin{tabularx}{\textwidth}{lll}
\textbf{Minimize:} & $\set{c}^T \set{x}$ & \hspace{1mm}Objective function \vspace{2mm}\\
\textbf{Subject to:} & $\set{A}\set{x} \leq \set{b}$ & \hspace{1mm}Constraints \vspace{1mm}\\
\end{tabularx}
\end{center}
A linear program consists of a set of $n$ continuous variables $\set{x} \in \vectorspace{R}{n}$, an associated set of costs for each variable $\set{c} \in \vectorspace{R}{n}$, and a set of $m$ linear constraints, denoted here by $\set{A} \in \vectorspace{R}{m \times n}$ and $\set{b} \in \vectorspace{R}{m}$.  The goal, or \emph{objective function}, is to find values for the set $\set{x}$ such that the total cost is provably minimized, whilst respecting the constraints.
In a MIP model, some or all of the variables are furthermore constrained to have integer values.

\begin{mipmodel*}
\begin{center}
\begin{tabularx}{\textwidth}{|llrXr|}
\hline
\textbf{Variables:}
& $x_{i} \in \{0, 1\}$	& $\forall i$	& Type of atom site $i$ (A=0, B=1)
& \reqnum\label{seq:cemodel_x}\Bstrut{2}\\
& $y_{\set{c}_f} \in \{0, 1\}$	& $\forall \set{c}_f$	& Cluster instance variable (off=0, on=1)
& \reqnum\label{seq:cemodel_ys}\Bstrut{3}\\
\textbf{Parameters:}
& $E_{\set{C}} \in \vectorspace{R}{}$	& $\forall \set{C}$	& Energy of cluster $\set{C}$
& \reqnum\label{seq:cemodel_bond_energies}\Bstrut{2}\\
& $N_B \in \vectorspace{N}{}$	& & Number of B-type atoms
& \reqnum\label{seq:cemodel_numb}\Bstrut{3}\\
\textbf{Minimize:}
& $\sum\limits_{\set{C}} \sum\limits_{\set{c}_f \in \set{C}}  E_{\set{C}} y_{\set{c}_f}$ &
& Total energy of system & \reqnum\label{seq:cemodel_objective}\Bstrut{3}\\
\textbf{Subject to:}
& $\sum\limits_{i} x_{i} = N_B$ &
& Fixed number of B-type atoms & \reqnum\label{seq:cemodel_numsolute}\Bstrut{2}\\
& $y_{\set{c}_f} \leq x_i$ & $\forall i \in \set{c}_f$ & Any atom absent from $\set{c}_f$ $\Rightarrow$ $y_{\set{c}_f}$ off
& \reqnum\label{seq:cemodel_bonds1}\Bstrut{2}\\
& $y_{\set{c}_f} \geq 1 - |\set{c}_f| + \sum\limits_{i \in \set{c}_f} x_i$ & & All atoms present in  $\set{c}_f$ $\Rightarrow$ $y_{\set{c}_f}$ on
& \reqnum\label{seq:cemodel_bonds2}\Bstrut{2}\\
\hline
\end{tabularx}
\end{center}
\caption{A MIP model for determining the configuration which provably minimizes the energy of a CE model.  The placement of the sites as well as the ECIs are fixed; the model determines the optimal chemical ordering only.  For any cluster instance with a positive [negative] ECI, the constraint given by Equation~(S\ref{seq:cemodel_bonds1}) [Equation~(S\ref{seq:cemodel_bonds2})] is redundant.}
\label{model:single_element}
\end{mipmodel*}

Scheme~\ref{model:single_element} shows the MIP model for determining the ground state chemical ordering of a bimetallic nanoparticle.  Each predetermined site, with index $i$, has an associated binary variable $x_i$~(S\ref{seq:cemodel_x}) which determines whether an A-type or B-type atom is placed at that site.  The system can (optionally) be constrained to contain $N_B$ B-type atoms, using Equation~(S\ref{seq:cemodel_numsolute}).  The activity of a cluster instance, indicated by a binary variable $y_{\set{c}_f}$~(S\ref{seq:cemodel_ys}), is governed by Equations~(S\ref{seq:cemodel_bonds1}) and (S\ref{seq:cemodel_bonds2}); taken together, these constraints are equivalent to the relation $y_{\set{c}_f} = \prod_{i \in \set{c}_f} x_i$.
Lastly, associated with each cluster is a predetermined ECI~(S\ref{seq:cemodel_bond_energies}) which is used to determine the total energy of the system~(S\ref{seq:cemodel_objective}).  Thus, the objective of the model is to choose how to order the A-type and B-type atoms such that the total energy of the system is minimized.

In addition to ground state configurations, a MIP model can be extended to find configurations of higher energy by adding linear constraints which forbid specific solutions, known as \emph{set-covering constraints}~\cite{chandru2011optimization}.
Given a known ground state configuration, $\set{x}^\prime$, we can forbid it with the constraint:
\begin{equation}
\sum\limits_{i \;:\; x_i^\prime = 0} x_i + \sum\limits_{i \;:\ x_i^\prime = 1} \left(1 - x_i\right) \geq 1
\label{eq:cluster_forbid}
\end{equation}
Constraints of this form are added for all solutions which are symmetrically equivalent to $\set{x}^\prime$.
When re-solving the MIP model, the ground state solution is now forbidden, and the lowest excited state is found instead.

The MIP method we apply is a widely-used technique in applied mathematics, logistics and industrial planning, but has not previously been used for the solution of ground states and excited states in a CE model.
A different approach to finding provably optimal ground states was
recently demonstrated by Huang et al.~\cite{huang2016maxsat}, making
use of pseudo-Boolean optimization rather than MIP, and applying it to
bulk alloys.

\subsection*{Cluster Selection and ECI Fitting}
For a system as large as a 309-atom nanoparticle, full electronic structure energy calculations are very time consuming.  As such, sampling a sufficient number of configurations for a CE model is impractical.
Instead, we use the semi-empirical Effective Medium Theory (EMT)~\cite{jacobsen1996semi} potential to calculate energies.  Semi-empirical potentials are fast to evaluate and highly accurate in bulk systems, but typically mispredict the energies of surface atoms.  Nonetheless, whilst the energies might be quantitatively inaccurate, the different site types in a 309-atom nanoparticle have such great variation in energy that we can expect the energy difference of different configurations to be at least qualitatively correct.

To fit the ECI parameters, we have sampled 75,000 chemical ordering configurations at a range of high and low energies (c.f.~Figure~\ref{fig:cluster_hull}).  The energy of each configuration has been minimized using gradient
descent to allow local relaxations without changing the overall structure of the nanoparticle.  The constructed CE model, which contains 60 1-body, 2-body and 3-body clusters, has a root-mean-square error (RMSE) of $0.060$ meV/atom.  Full details on the CE model can be found in the supplementary information section S3.

The approach described here is not specific to 309 atom Mackay icosahedra - it generalizes well to other nanoparticle morphologies.  A precondition for the use of a CE model, however, is the ability to identify a set of atomic sites.  Given sufficient training data the energetic effects of relaxations can be captured in the CE model, given that the relaxations are of limited magnitude and strictly local i.e. they do not affect the neighbour relationships between atomic sites.
Whilst the solution time of a MIP model is dependent on many factors, our experience with the nanoparticle described here suggests that solving systems larger than a few hundred atoms is prohibitively expensive without reducing the number of clusters, and thereby the accuracy of the CE model.

\begin{acknowledgments}
PML~thanks John Connelly for productive discussions, Christopher Schuh for an  introduction to the nanoparticle design problem, and Jesper Larsen for advice on selecting solver settings.  This work was supported by
by research grants 7026-00126B and 1335-00027B from the Danish Council for Independent Research and grant 9455 from VILLUM FONDEN.
\end{acknowledgments}

\bibliography{combinedrefs}

\end{document}